**Title Page:**

Title: Female and Combined Male-Female Injury Risk Functions for the Anterior Pelvis Under Frontal Lap Belt Loading Conditions

Corresponding Author: Connor Hanggi[1]

4040 Lewis and Clark Drive, Charlottesville, VA, 22911

zwk8nj@virginia.edu

ORCID: 0009-0002-8779-3914

Contributing Authors: Joon Seok Kong[1], James Caldwell[1], Bronislaw Gepner[1], Martin Östling[2], Jason R. Kerrigan[1]

1: Center for Applied Biomechanics, University of Virginia

4040 Lewis and Clark Drive, Charlottesville, VA, 22911, United States

2: Autoliv Research

Wallentinsvägen 22, SE-447 83, Vårgårda, Sweden

Abbreviated Title: Injury Risk Function for the Female Anterior Pelvis

**Acknowledgements**: The authors would like to acknowledge Autoliv Research for providing financial and technical support for this study, as well as the tissue donors for their generosity, which made this study possible. The authors would like to thank everyone at the University of Virginia Center for Applied Biomechanics who assisted with this project, with special acknowledgement to Jeesoo Shin, Brian Overby, Joey White, Brandon Hale, Sara Sochor, and Pavel Chernyavskiy.

**Author Contribution Declaration**:

All authors contributed to the study conception and design. Material preparation, data collection and analysis were performed by Connor Hanggi, James Caldwell, and Joon Seok Kong. The first draft of the manuscript was written by Connor Hanggi and all authors commented on previous versions of the manuscript. All authors read and approved the final manuscript.

**Abstract**:

Purpose: Iliac wing fractures due to lap belt loading have been observed in laboratory settings for 50 years and recent data suggest they are also occurring in the field. Automated driving systems (ADS) and other occupant compartment advancements are expected to offer enhanced flexibility in seating orientation, which could place a greater reliance on the seatbelt to restrain occupants. Such changes may increase seatbelt loads and create new challenges in successfully restraining occupants and mitigating injury to areas such as the pelvis. Injury criteria exist for component-level male iliac wing fractures resulting from frontal lap belt loading, but not for females.

Methods: This study explored female iliac wing fracture tolerance in the same loading environment as a previous study that explored the fracture tolerance of isolated male iliac wings. Male and female fracture data were combined to evaluate the effect of sex. Injury risk functions were created by fitting Weibull survival models to data that integrated censored and exact failure observations.

Results: Twenty female iliac wings were tested; fourteen of them sustained fracture with known failure forces (exact), but the remaining six wings either (1) did not fracture, or (2) fractured after an event that changed the boundary conditions (right censored). The fracture tolerance of the tested specimens ranged widely (1134 – 8759 N) and averaged 4240 N (SD 2516 N).

Conclusion: Female data and combined male-female data were analyzed. Age was the only covariate investigated in this study that had a statistically significant effect and improved the predictive performance of the models.

**Introduction**:

The introduction of automated driving systems (ADS) has led the automobile industry to predict that occupants will desire more comfortable seating positions during the ride such as forward-facing postures with reclined seatbacks [1-3] and/or rearward positioned seats with more legroom [4]. Such seating positions may present new challenges in how to successfully restrain occupants. For example, reclined occupants in frontal crashes tend to exhibit high pelvis excursion and, as a result, are subjected to substantial lap-belt forces if no knee bolster is present [5].

In 2019, sled tests were conducted on five male post-mortem human surrogates (PMHS) [6]. In this test series, no knee bolster was present, and two of the five PMHS sustained iliac wing injuries on the outboard side between the anterior superior iliac spine (ASIS) and anterior inferior iliac spine (AIIS) (Supplementary Material Fig. A1). Peak lap belt forces for these two subjects were 6.6 kN and 7.8 kN. In 2023, the same sled tests were conducted on 3 female PMHS. The peak lap belt loads for the two female fracture cases were 5.8 kN and 6.9 kN [7].

Further investigation of the literature showed that similar iliac wing fractures occurred in several other frontal sled test series where no knee bolster was present [8-10]. In fact, the iliac wing fractures due to lap belt loading have been noted to occur in full vehicle test environments (i.e., even with a knee bolster) since the 1970s [11-12]. Seeking to address this problem, recent studies have provided insight on pelvis fracture tolerance in male PMHS [13-14]. However, no injury tolerance data is available for the female pelvis.

Similar injuries have also been noted in field data. A Crash Injury Research and Engineering Network (CIREN) database (years 2004–2020) query identified 56 frontal crashes with the Abbreviated Injury Scale (AIS) code (856151.2), which indicates a pelvic fracture with an intact posterior arch. Further investigation showed that 14 cases were identified where occupants sustained lap belt-induced iliac wing fractures [14].

Despite the occurrence of these injuries, no experimental studies had been devoted to studying the cause or tolerance of such iliac wing injuries until 2021. In 2021, an initial experimental study was completed [13] where a simplified loading environment ("belt pull") was created to isolate the instance of lap belt loading that caused fracture in sled tests described by Richardson et al. [6]. Stationary lap belt loading experiments were conducted to remove the inertial effects present in the sled environment and focus on lap belt loading as a mechanism of injury. Lap belt-induced iliac wing fracture occurred in two PMHS with approximately 90° lap belt-to-pelvis notch angles. The lap belt-to-pelvis notch angle (Supplementary Material Fig. A2) was defined as the angle between the long axis of the belt and the line connecting the ASIS and the AIIS [13]. The size, location, and complexity of the fractures in those two PMHS were similar to those seen in the sled tests described by Luet et al. [8], Uriot et al. [9], and Richardson et al. [6].

Recently, another study was conducted which used a further simplified environment to generate fractures in denuded iliac wings via direct belt webbing loading. The authors [14] measured the fracture tolerance of twenty two male PMHS iliac wing specimens and created an injury risk function that estimates risks for male occupants. Simplifying the lap belt-to-pelvis loading environment permitted direct measurement of the force applied to the iliac wing at fracture, which can be related to a measurable quantity of lap belt load [15] and thus permits the application of the resulting injury risk function to computational human body models (HBMs) and anthropomorphic test devices (ATDs). Moreau et al. [14] followed typical approaches used in injury biomechanics to develop injury risk functions using parametric Weibull survival models [16-18], although additional analyses were also discussed.

While pelvis fracture tolerance data and injury risk curves for male PMHS provide a strong basis for guiding restraint design, the findings cannot be assumed to apply to the female population. Previous research has shown differences in pelvis geometry and size between males and females [19]. Various other studies have indicated that sex has a significant effect on injury in automotive crashes [20-23]. In fact, a 2019 study analyzing automobile injury trends for belted occupants in frontal collisions found that females exhibited a greater risk of lower extremity injuries, even after controlling for delta-V, age, height, BMI, and vehicle model year [20]. The increased risk is likely due to a combination of tolerance and exposure, but it illustrates the importance of evaluating tolerance. The authors are unaware of other research that investigates the differences between male and female iliac wing fracture tolerance. Therefore, this study focuses on measuring fracture tolerance data for female iliac wings and comparing it to male iliac wing fracture tolerance. Fracture tolerance data for female pelvises is crucial for ensuring safe and effective restraint systems.

The goal of the current study was to replicate the test methods in Moreau et al. [14] using female PMHS iliac wing specimens to measure each specimen's fracture tolerance and create an injury risk function that estimates risks for the female population. The data gathered from this study was then combined with the male PMHS iliac wing fracture data from Moreau et al. [14] to obtain a more representative portrayal of pelvis fracture tolerance across the entire population and to test whether sex affects fracture tolerance of the iliac wing.

**Materials and Methods**:

This study included ten female, fresh-frozen PMHS pelvises (Table 1). The specimens were obtained and treated in accordance with the ethical guidelines established by the Human Usage Review Panel of the National Highway Traffic Safety Administration, and all testing and handling procedures were reviewed and approved by an institutional review board at the University of Virginia (Charlottesville, VA, USA). The specimens were frozen until testing and screened for blood-borne pathogens including HIV and Hepatitis B and C. An initial computed tomography (CT) scan confirmed the absence of preexisting bone injury and abnormalities.

*Table 1: Specimen anthropometry and dimensions (PSIS: posterior superior iliac spine, PC: pubic crest).*

| Subject ID | Sex | Age | Weight (kg) | Stature (cm) | ASIS-PSIS distance L/R (mm) | | ASIS-PC distance L/R (mm) | |
|---|---|---|---|---|---|---|---|---|
| 1055 | F | 84 | 70.7 | 175.3 | 166 | 166 | 143 | 146 |
| 1075 | F | 66 | 63.5 | 167.6 | 175 | 173 | 146 | 137 |
| 1077 | F | 66 | 65.3 | 167.6 | 161 | 163 | 122 | 126 |
| 1078 | F | 68 | 41.7 | 167.6 | 159 | 165 | 138 | 150 |
| 1079 | F | 35 | 87.1 | 170.2 | 165 | 165 | 122 | 132 |
| 1080 | F | 23 | 51.7 | 172.7 | 157 | 159 | 128 | 127 |
| 1090 | F | 36 | 88.9 | 167.6 | 156 | 155 | 118 | 120 |
| 1091 | F | 40 | 36.3 | 157.5 | 158 | 165 | 117 | 126 |
| 1092 | F | 51 | 36.3 | 162.6 | 148 | 151 | 127 | 133 |
| 1094 | F | 44 | 65.8 | 175.3 | 161 | 161 | 122 | 123 |

*Specimen Preparation*

Surrounding tissue was removed from each pelvis and the sacroiliac joints and pubic symphysis were opened to separate the iliac wings from one another. The iliac wings were potted in the same fashion as done previously with male iliac wings, which aimed to minimize the roll angle of each iliac wing relative to the vertical sides of the potting cup (Supplementary Material Fig. A3) [14]. Each wing was additionally oriented so that the line connecting the ASIS and AIIS (the "Notch Plane") was as parallel to the top edge of the potting cup as possible. "Notch Angle" (the angle between a line connecting the ASIS and AIIS relative to a horizontal plane of the potting cup, defined by the plane connecting the threaded inserts inside of the potting material) measurements for each wing were completed in alignment with the male iliac wings (Supplementary Material Fig. A3). The Notch to Belt Angle target in this study was 90° for all tests, which replicated those from Moreau et al. [14] and Richardson et al. [6]. Moreau et al. [14] showed that belt loading angle (75-90 deg) was not a significant predictor of iliac wing fracture force, therefore this study investigated two different belt positions along the anterior iliac spine. Position 0 was the same placement used in Moreau et al. [14], where the upper edge of the belt webbing was aligned with the most anterosuperior point of the ASIS (Figure 1a). Position 50 was achieved by aligning the belt webbing midline with the ASIS (Figure 1b).

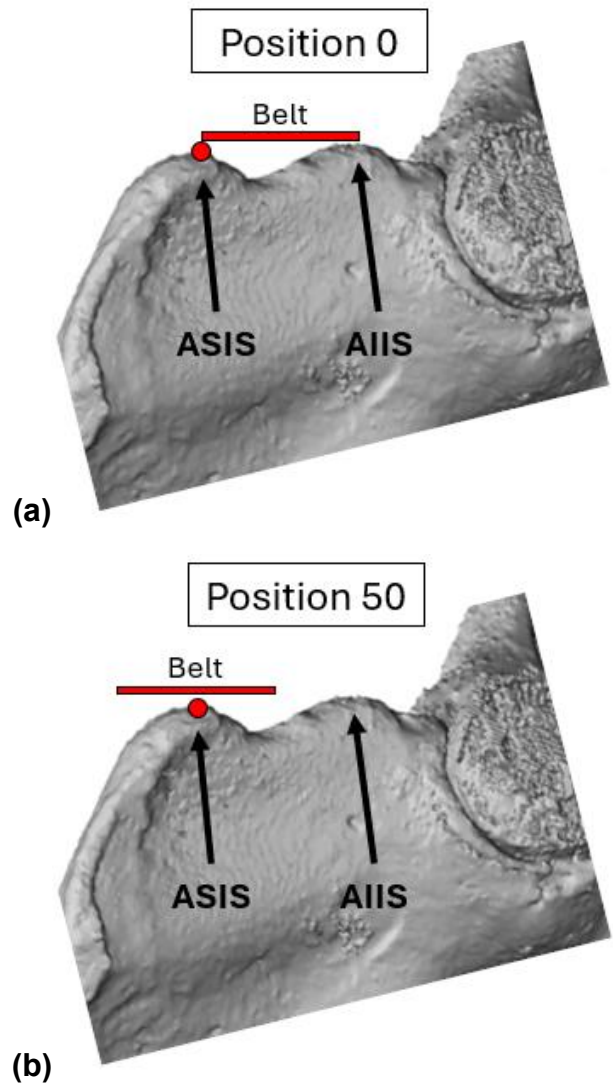

*Figure 1: **a** Belt Loading Position 0, and **b** Belt Loading Position 50.*

*Test Setup and Instrumentation*

The belt fork fixture that was used to hold the belt webbing and load the iliac wing specimens consisted of two fork tines, which the belt webbing spanned across (Figure 2). Clamping plates on either side of the fixture were used to secure the belt webbing and tensioning bolts were threaded through the clamping plates. As the bolts were tightened, the clamps would raise and pull on both sides of the belt webbing, thus tensioning the belt. Consistent belt tension between tests was achieved by raising the clamping plates to the same height each time the belt was tensioned for a new test. The clamping plates were designed with the expectation of bearing high loads, since Moreau et al. recorded a fracture force of greater than 9.5kN in one test and greater than 8kN in multiple tests [14]. The clamping plates had teeth to grip the belt webbing and minimize the potential for slipping. The bottom clamping plate was wide enough to close the gap with the fork tine and thus eliminate internal bending of the tensioning bolts. The clamping plates were made of steel and thickened to minimize plate bending. While the belt webbing was not observed to slip through the clamps in Moreau et al. [14], high loads were observed and expected for the current test series. Therefore, the fixture was modified from Moreau et al. [14] to improve grip on the belt and ensure the webbing did not slip through the fixture at high loads.

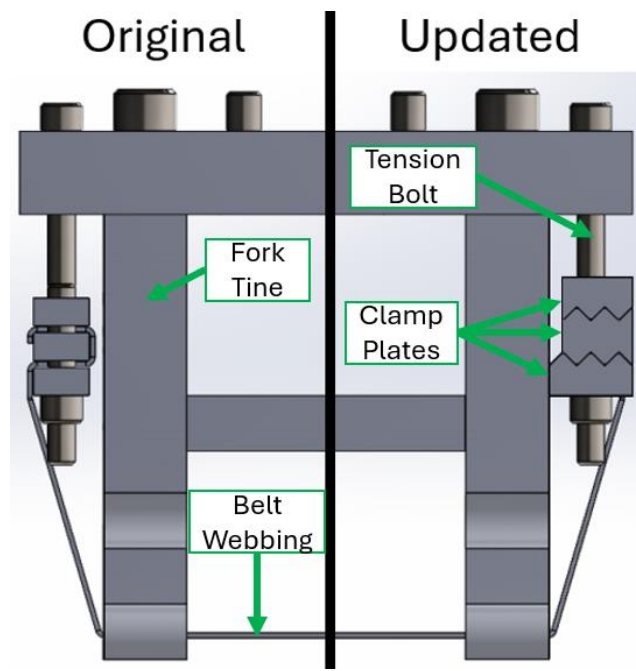

Figure 2: Belt fork clamp designs for original study [14] (left) and current study (right).

Loadcells (one six-axis loadcell and two single-axis loadcells) below the potting cup, a strain gauge rosette (Micro-Measurements ® C4A-06-062WW-350-39P) on the lateral iliac wing, and a string potentiometer on the belt fork fixture were all placed in accordance with the test setup in Moreau et al. [14]. Loadcells recorded the reaction forces, the strain gauge rosette assisted in determining fracture timing, and the string potentiometer recorded the displacement time history. Sensor data were not filtered.

*Test Methods*

Wing fracture tests were performed on all 20 female iliac wing specimens. Using measurements from a pre-test CT analysis, each wing was mounted to the testing machine (Supplementary Material Fig. A4) and oriented to achieve the desired 90°

Notch to Belt Angle target to match Moreau et al. [14]. Each wing was randomly assigned a belt loading position, with left and right wings from the same donor being assigned different loading positions. The entire potting cup was translated accordingly to place the belt in Position 0 or Position 50. The belt fork was then displaced downward until a ~10 N preload was applied to ensure that the belt was in contact with the pelvis surface. The actuator of the biaxial machine was then displaced vertically downward at a rate of ~1 m/s for a displacement of 15–43 mm, which was dependent upon fixture and anatomic displacement limits. Exact displacement targets were determined for each specimen by determining the maximum stroke that could occur before the metal belt fork fixture interacted with the wing or the potting cup.

Two wings for each of the ten female PMHS were tested, which produced 20 observations. Force data were either recorded exactly and deemed uncensored, or they were right censored. Uncensored fracture forces were recorded in cases where the exact time and force of fracture were identified using high-speed video review or by identifying abrupt changes in the reaction forces or pelvis surface strains. Fracture forces were considered right-censored data points in one of two scenarios: (1) no fracture occurred after the maximum allowable loading displacement, or (2) fracture occurred after an event that caused the boundary conditions to change (e.g. belt webbing tear). In case (2), the maximum force prior to a change in boundary conditions was recorded as the censored force value.

*Data Analysis*

Weibull survival (or time-to-event) analysis, which can accommodate data censoring, was used to model the force at fracture. The procedure followed that of Moreau et al. [14] and is further described in McMurry and Poplin [17], as it is commonly used in the biomechanics field to handle double censored data [16-17]. Using the female fracture data, a base model with no covariates and a univariate model incorporating belt position, as well as their associated risk curves, were created. Univariate survival models were created separately for the male and female datasets to examine each of the following covariates: age, weight, height, ASIS-to-PSIS distance, and ASIS-to-pubic crest (PC) distance.

Additional survival models were created from combined male and female fracture data, which included a base model with no covariates and univariate models with age, sex, belt angle, and belt position covariates. Risk curves were plotted for the univariate sex and age survival models using the combined dataset.

Injury risk functions were compared using the Corrected Akaike Information Criterion (AICc) [24], which is particularly well suited to small samples. All statistical analyses were completed in R version 4.3.1, using the "survival" package [25].

**Results**:

*Fracture*

Fourteen out of twenty iliac wings sustained fracture with known failure forces and were exact-censored (Table 2). The remaining six wings were right-censored. There were no left-censored tests. Right censoring was necessary in four tests (1075R, 1079L, 1091R, 1094L) because the belt webbing tore before the pelvis fractured, which changed the boundary conditions of the loading (the belt-to-pelvis loading area abruptly changed). The iliac wings in the other two right-censored tests did not sustain fracture. One wing (1079R) did not fracture due to anatomical limitations on the displacement input. The other test (1080R) did not fracture, but by the time the testing machine reached 9096 N, the displacement rate lowered substantially, so the test was right censored at this time (the decision was made due to the rate sensitivity of bone [26]). Forces for the uncensored data points ranged from 1135 to 8759 N (mean 4240 N, SD 2516 N).

*Table 2: Loading tolerance data.*

| Subject | Force at Event (N) | Censoring | Belt Position |
|---|---|---|---|
| 1055L | 4300 | Exact | 50 |
| 1055R | 4864 | Exact | 0 |
| 1075L | 6418 | Exact | 0 |
| 1075R | 4857 | Right | 50 |
| 1077L | 2091 | Exact | 0 |
| 1077R | 1665 | Exact | 50 |
| 1078L | 1135 | Exact | 50 |
| 1078R | 1688 | Exact | 0 |
| 1079L | 7579 | Right | 50 |
| 1079R | 4630 | Right | 0 |
| 1080L | 8759 | Exact | 0 |
| 1080R | 9096 | Right | 50 |
| 1090L | 2747 | Exact | 0 |
| 1090R | 3369 | Exact | 50 |
| 1091L | 7636 | Exact | 0 |
| 1091R | 5783 | Right | 50 |
| 1092L | 2154 | Exact | 50 |
| 1092R | 5201 | Exact | 0 |
| 1094L | 4222 | Right | 50 |
| 1094R | 7330 | Exact | 0 |

*Survival Models*

The univariate survival models for height, weight, pelvis size, and belt position were not statistically significant in any dataset (Supplementary Material, Table A1). Injury risk curves were constructed for each belt position for comparison (Table 3, Figure 3b). The univariate survival model of the female dataset that incorporated age showed a statistically significant relationship (p<0.05) and had the lowest AICc value (Supplementary Material, Table A1).

When male and female data were combined to create a univariate survival model with sex as the covariate, sex did not display a statistically significant relationship with iliac wing fracture force (Supplementary Material, Table A1). An IRF was created from the survival model that incorporated sex (Table 3, Figure 3c). Results from combining male and female fracture data and creating univariate survival models with the covariates age, belt angle, and belt position showed that the age covariate was statistically significant, and the age univariate model had the lowest AICc value (Supplementary Material, Table A1). An IRF with age as the covariate was also made using the combined male-female data (Table 3, Figure 3d).

*Table 3: Model parameters with goodness of fit.*

| **Female Injury Risk Functions** | | | | | | |
|---|---|---|---|---|---|---|
| Model | AICc | Shape | Intercept | Position | Age | Injury Risk Function Pr(Fracture \| Force) |
| Base | 274.9 | 1.754 | 8.762 [8.464, 9.061] | [–] | [–] | $1 - e^{-(Force \times e^{-(8.762)})^{1.754}}$ |
| *Position* | 277.2 | 1.743 | 8.676 [8.299, 9.052] | 0.216 [-0.422, 0.855] | [–] | $1 - e^{-(Force \times e^{-(8.676 + .216 \times Position)})^{1.743}}$ |
| **Age** | 271.0 | 2.121 | 9.577 [8.851, 10.303] | [–] | -0.017 [-0.029, -0.005] | $1 - e^{-(Force \times e^{-(9.577 - .017 \times Age)})^{2.121}}$ |
| **Combined Male-Female Injury Risk Functions** | | | | | | |
| Model | AICc | Shape | Intercept | Sex | Age | Injury Risk Function Pr(Fracture \| Force) |
| Base | 598.8 | 1.682 | 8.698 [8.493, 8.902] | [–] | [–] | $1 - e^{-(Force \times e^{-(8.698)})^{1.682}}$ |
| *Sex* | 600.8 | 1.674 | 8.764 [8.451, 9.077] | -0.122 [-0.540, 0.296] | [–] | $1 - e^{-(Force \times e^{-(8.764 - 0.122 \times Sex)})^{1.674}}$ |
| **Age** | 591.4 | 1.877 | 9.810 [9.023, 10.598] | [–] | -0.020 [-0.032, -0.007] | $1 - e^{-(Force \times e^{-(9.810 - 0.020 \times Age)})^{1.877}}$ |

*Italicized survival model covariates were not significant.*
**Bold survival model covariates were significant and provided the best model fit.**

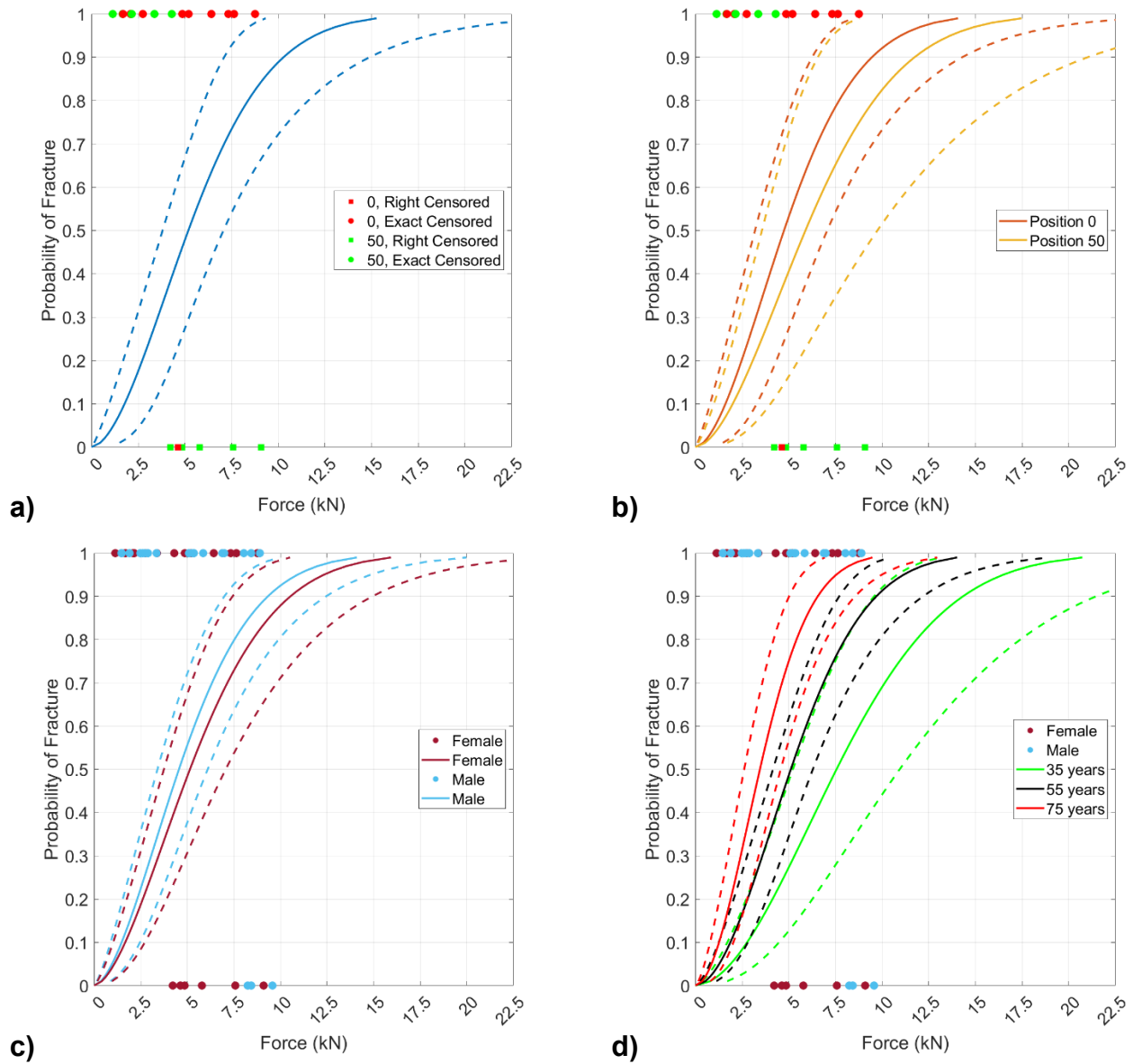

*Figure 3: **a** Base injury risk function for female fracture data (n=20) with 95% confidence interval, **b** injury risk function for female fracture data (n=20) with Position covariate and 95% confidence intervals, **c** injury risk function for combined male-female fracture data (n=42) with Sex covariate and 95% confidence intervals, **d** injury risk function for combined male-female fracture data (n=42) with Age covariate and 95% confidence intervals.*

**Discussion**:

This study is, to our knowledge, the first to investigate the injury tolerance of isolated, female iliac wings to loading similar to what is experienced by belted occupants in frontal crashes. The risk functions and associated confidence intervals developed and presented here utilize the iliac wing fracture tolerance, and that tolerance can be converted to belt force [15], which can be used to predict the risk of iliac wing fracture in ATDs and HBMs. With proper instrumentation, the force applied to the anterior iliac spine region of the iliac wing (between the ASIS and AIIS) in an experiment or simulation can be measured and compared to the risk functions to describe the risk of a fracture like those seen in this study.

Seat belt webbing that was stretched between tines of the loading fork was used to deliver the distributed load to the iliac wings, but it was not intended to directly represent the same loading applied by a seat belt through the abdominal soft tissue for belted occupants in frontal crashes. The seat belt webbing permitted the application of a distributed load to the iliac wing, which Moreau et al. [14] demonstrated to be the best representation of the boundary conditions and fracture types that were seen in belted humans in frontal crashes. Further, the use of seat belt webbing as the loading device permits subsequent studies to replicate the boundary conditions of these experiments in finite element modeling since detailed, validated belt webbing finite element models are commonly available.

*Fracture*

The fracture patterns of the female pelvises in this study (Supplementary Material, Figure A5) were similar to those seen in Moreau et al. [14] and were additionally observed in sled tests [6], [9]. There was substantial variability in the fracture tolerance across the samples included in this study (Table 2), and the range is comparable to the range of the male iliac wing fracture experiments [14]. Some wings fractured at less than 2000N, while others withstood over 8500N.

*Survival Models*

Unlike a correlation or regression analysis, Weibull survival models appropriately handle censored data and therefore can include the entire dataset in the analysis. Univariate Weibull survival models of the female fracture data supported findings from the male fracture data that weight, height, and pelvis size (both ASIS-to-PSIS and ASIS-to-PC) were not significant predictors of fracture force (Supplementary Material Table A1). The confidence interval and AICc value of the survival model with the belt position covariate indicated that including belt loading position did not improve predictive performance from the base model and added unnecessary complexity. Therefore, it is not beneficial to use belt position to predict fracture tolerance of the iliac wing (Table 3,

Figure 3b). In other words, we did not identify a statistically significant difference in the fracture tolerance when we centered the loading on the ASIS versus when we aligned the upper edge of the loading at the ASIS.

Using AICc as a performance metric, the univariate models with age best fit the data in both the female and combined male-female datasets (Table 3). However, in the male dataset, the AICc of the univariate model with age was lower than the AICc of the Base model by less than two units [27], so the two models are essentially equally supported by the data and there is no strong evidence to favor one model over the other (Supplementary Material, Table A1). The age range of the female dataset (23-84 years) was larger than the age range of the male dataset (50-77 years), which may help to explain this difference. Considering the age ranges, the statistical significance of the age covariate in all datasets, and the improvement in model fit in the female and combined datasets, it is likely that age is useful in predicting pelvis fracture tolerance. Widening the age range of the male dataset to include young males may be helpful in strengthening this notion.

When analyzing the male and female iliac wing fracture data together, the IRF indicated that using sex as a covariate was not statistically significant and did not improve predictive performance of the model (Table 3, Figure 3c). We could not identify a sex-effect in the fracture tolerance of the iliac wing under frontal lap belt loading, which suggests it is not beneficial to use sex as a covariate to predict iliac wing fracture tolerance.

*Bone Density*

Moreau et al. [14] showed that bone density (BD) and fracture tolerance in male iliac wing specimens were correlated and that BD improved the predictive capability of the injury risk function. The BD measurement that was used by Moreau et al. [14] was a weighted average calculation of BD for the entire iliac wing, which includes both cortical and trabecular bone, as well as areas of the pelvis that are further away from the anterior iliac spine region that was loaded. Though not part of the current study, microstructural bone measurements of the ASIS and AIIS for each tested male and female iliac wing were collected via MicroCT. Rather than using the weighted average BD calculation that was used by Moreau et al. [14], future work will examine local bone microstructural properties, such as cortical thickness and trabecular density, at the fracture sites to determine if they are significant predictors of iliac wing fracture tolerance.

*Limitations*

Ideally, this study would have resulted in twenty independent samples for statistical analysis. However, the effective sample size of the study is somewhere between ten and twenty due to the correlation seen between the left and right sides of the same pelvis (Supplementary Material Fig. A6). As was the case for Moreau et al. [14], our statistical models assume all observations are mutually independent, which may bias the statistical model. Further methodological extensions are necessary to quantify (1) the degree of clustering of failure forces in the context of injury risk functions; and (2) estimation, interpretation, and validation of models that allow for correlation between observations.

Due to the complex shape of the pelvis, potting all iliac wing specimens the exact same way was not possible. The test fixture allowed us to successfully account for the difference in the Notch plane angle by rotating the potting cup on the testing machine; however, we could not adjust the roll angle of the potting cup. In addition, subject variability in pelvis shape made it nearly impossible to control differences in the wing geometry that would interact with the belt fork tine and limit the amount of displacement available. Similarly, some wings had more pronounced ASIS landmarks, which may have influenced belt webbing tears. A more prominent ASIS is less likely to support distributed loading on the superior half of the belt webbing, and therefore allows a transverse tension force in the belt that may be less common in sled tests and real frontal crashes due to the layers of soft tissue that help distribute the load.

*Conclusions*

The injury tolerance data collected as a part of this study are the first for female iliac wings under frontal belt loading conditions. The testing procedure outlined in this paper was successful in replicating that of Moreau et al. [14] for female iliac wing specimens. Accordingly, we generated fracture at the desired location and of the desired type and severity. Fourteen out of twenty tested female iliac wings fractured, all of which represented injuries seen in the literature that occurred due to lap belt loading of the pelvis. Sex did not display a significant relationship with iliac wing fracture tolerance. Of all the survival models created in this study, age was the only covariate that improved the prediction in the female and combined datasets. In the male dataset, however, the weighted average bone density calculation was the best predictor, suggesting that some aspect about the bone quality may be able to better explain the large variation in iliac wing fracture tolerance. Localized anterior iliac spine microstructure measurements and their connection with fracture tolerance will be investigated in the future.

**Declarations**:

*Funding*

This study was funded by Autoliv Research (Vårgårda, Sweden).

*Competing Interests*

The authors have no competing interests to declare that are relevant to the content of this article.

**Supplementary Material for**

Hanggi C, Kong J, Caldwell J, Gepner B, Östling M, Kerrigan J. *Female and Combined Male-Female Injury Risk Functions for the Anterior Pelvis Under Frontal Lap Belt Loading Conditions*. Annals of Biomedical Engineering. (2025).

# Contents

Figure A1: 3D reconstructions of pelvis fractures from PMHS sled tests [6].

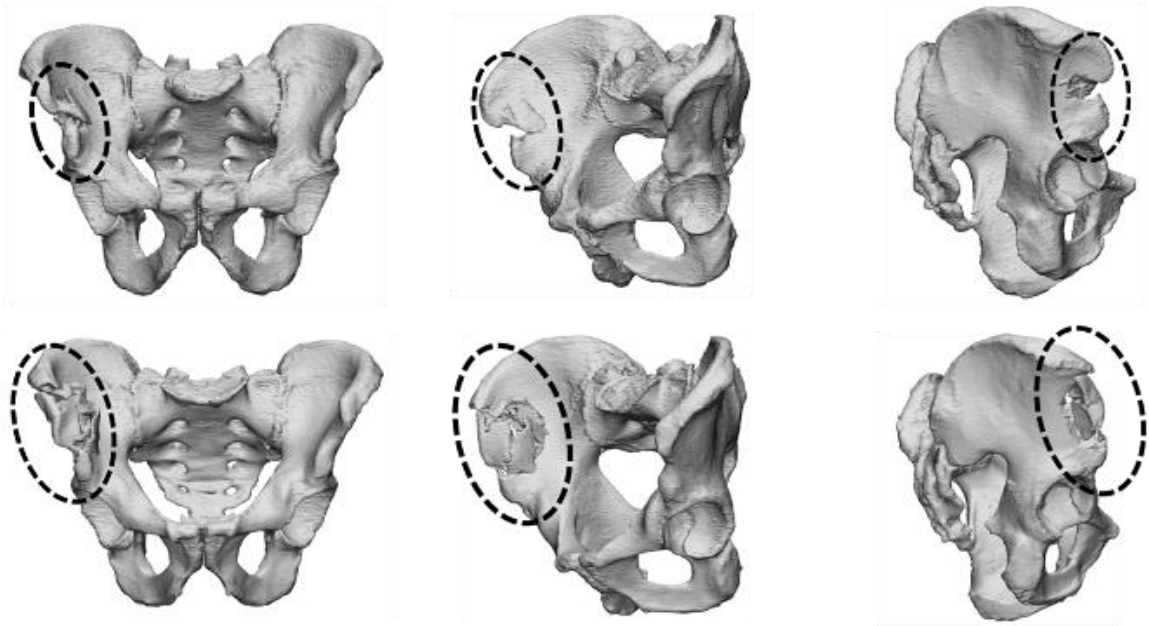

Figure A2: Notch to belt angle diagram.

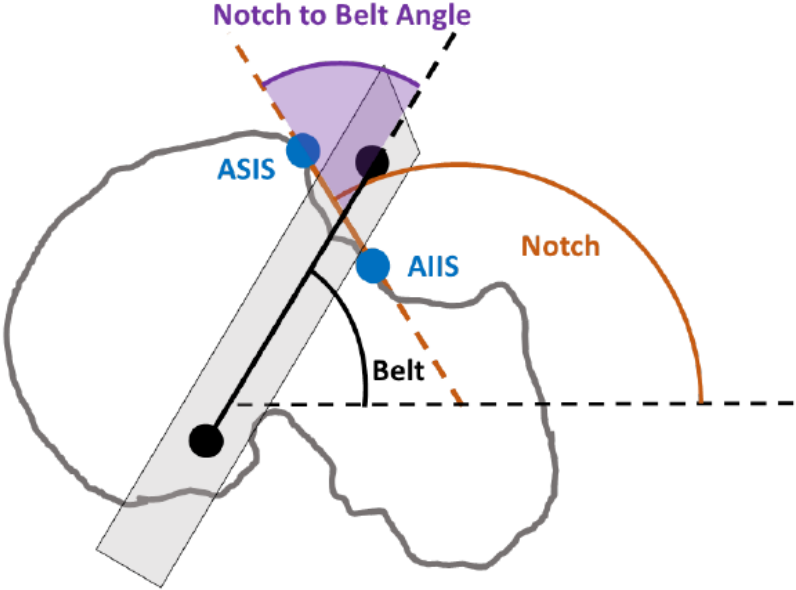

Figure A3: a) lateral/medial and b) superior/inferior views of a potted iliac wing illustrating the relationship between the wing and potting cup coordinate systems. The Notch to Belt Angle (green) is shown in c) as the difference between the Notch angle (purple) and the belt/loading angle (black). The Notch to Belt Angle was adjusted by rotating the iliac wing and potting cup on the test fixture within the plane shown in a). Adjustment of the pelvis roll angle was accomplished by rotating the iliac wing relative to the potting cup during the potting process in the plane shown in b).

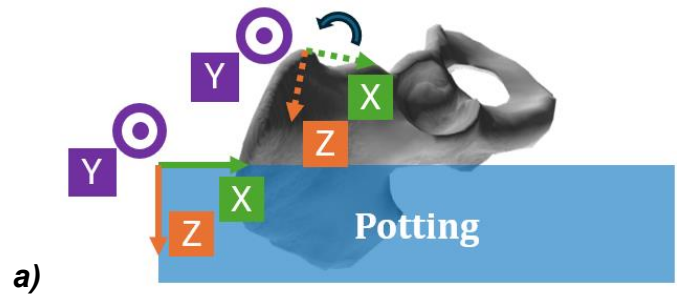

a)

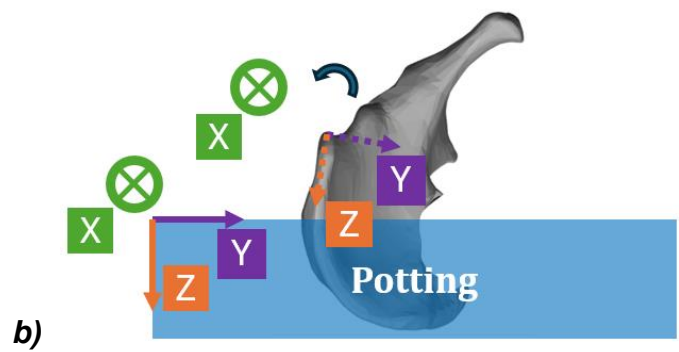

b)

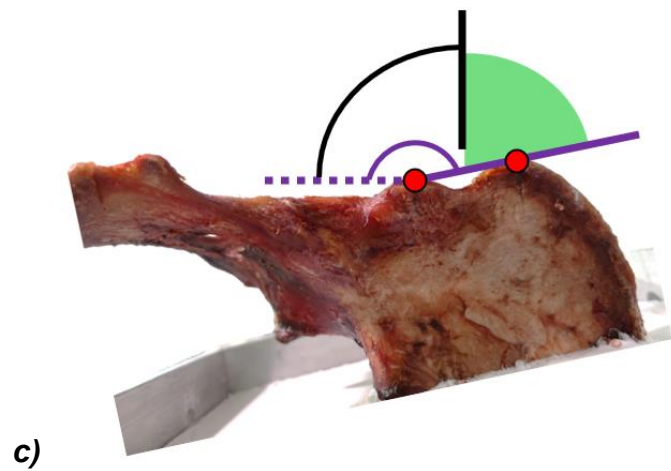

c)

Figure A4: CAD image of biaxial testing machine and test setup for component-level iliac wing fracture experiments.

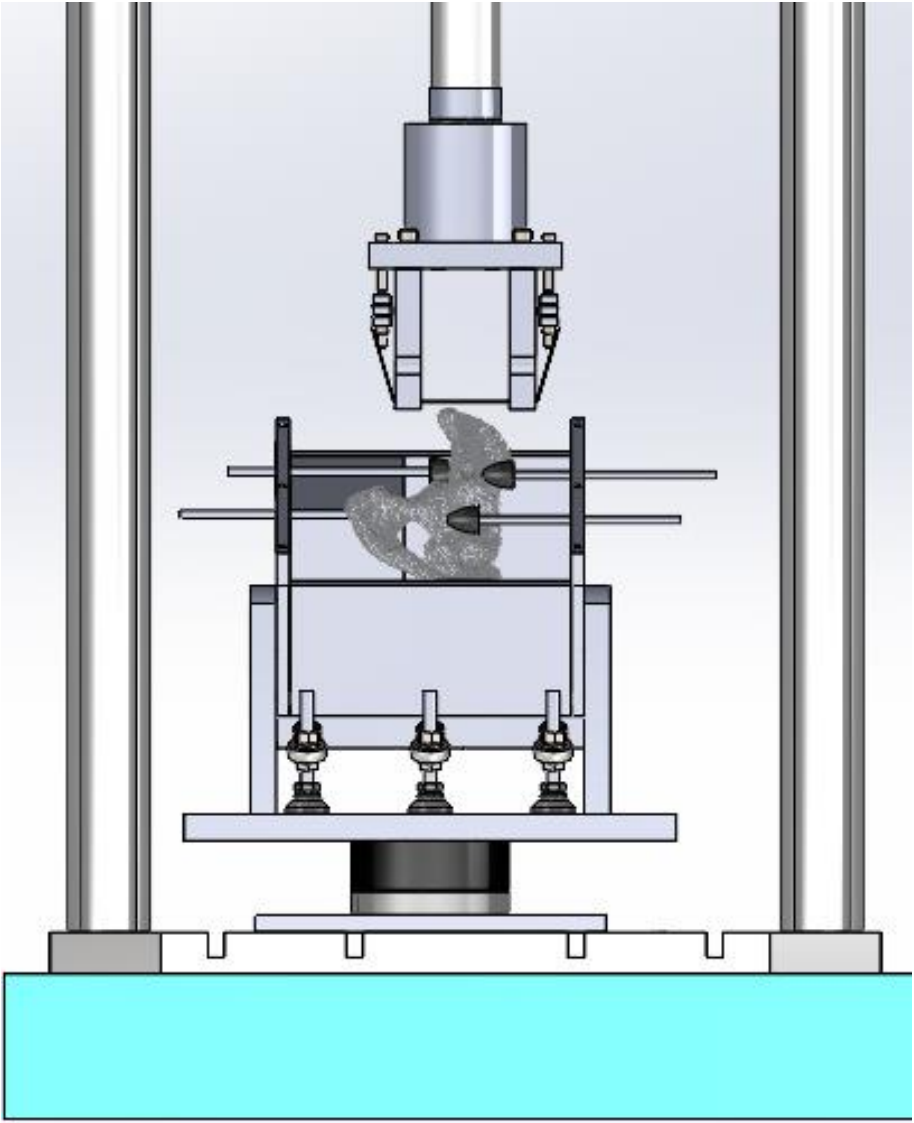

Table A1: Univariate Weibull survival model AICc values and p-values for each dataset (male, female, combined).

| Model | AICc | p-value (covariate) |
|---|---|---|
| **Male** | | |
| Base | 328.6 | [–] |
| BD | 324.5 | 0.001* |
| Age | 327.1 | 0.033* |
| Height | 328.3 | 0.084 |
| Weight | 331.2 | 0.741 |
| ASIS-PSIS | 329.9 | 0.234 |
| ASIS-PC | 330.0 | 0.250 |
| Angle | 331.1 | 0.683 |
| **Female** | | |
| Base | 274.9 | [–] |
| Age | 271.0 | 0.007* |
| Height | 277.4 | 0.627 |
| Weight | 277.7 | 0.887 |
| ASIS-PSIS | 276.5 | 0.306 |
| ASIS-PC | 276.5 | 0.259 |
| Position | 277.2 | 0.506 |
| **Combined** | | |
| Base | 598.8 | [–] |
| Age | 591.4 | 0.003* |
| Sex | 600.8 | 0.567 |
| Position | 600.4 | 0.399 |
| Angle | 600.7 | 0.500 |

Figure A5: Example iliac wing injury from test (specimen 1078R).

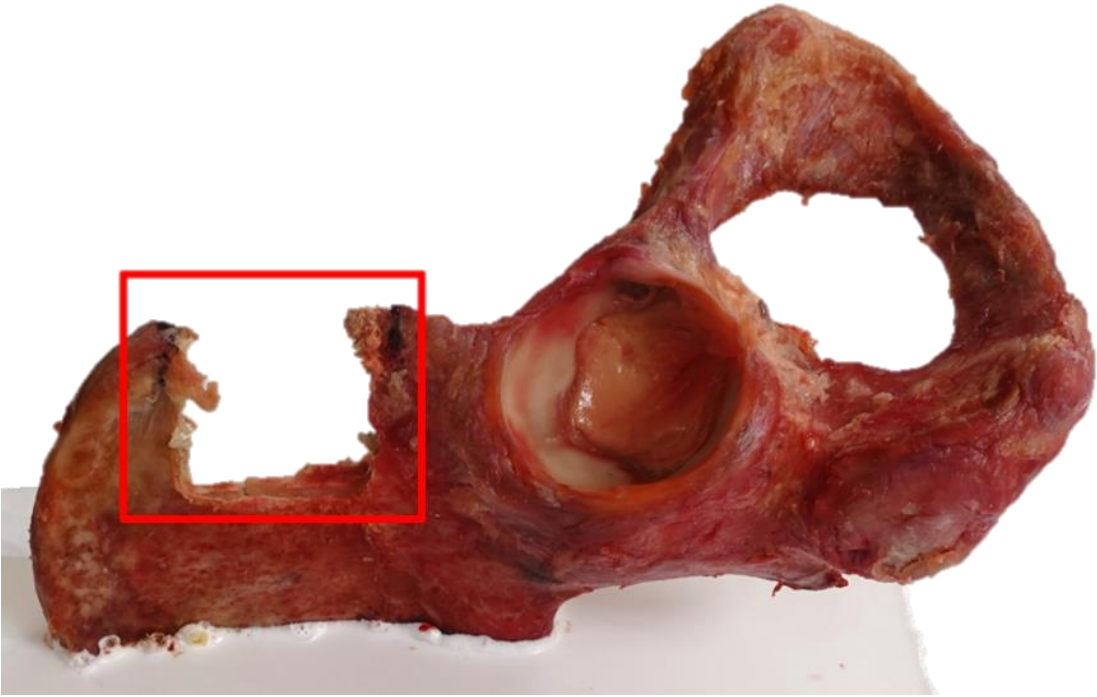

Figure A6: Correlation plot of fracture force for left and right sides of each female pelvis.

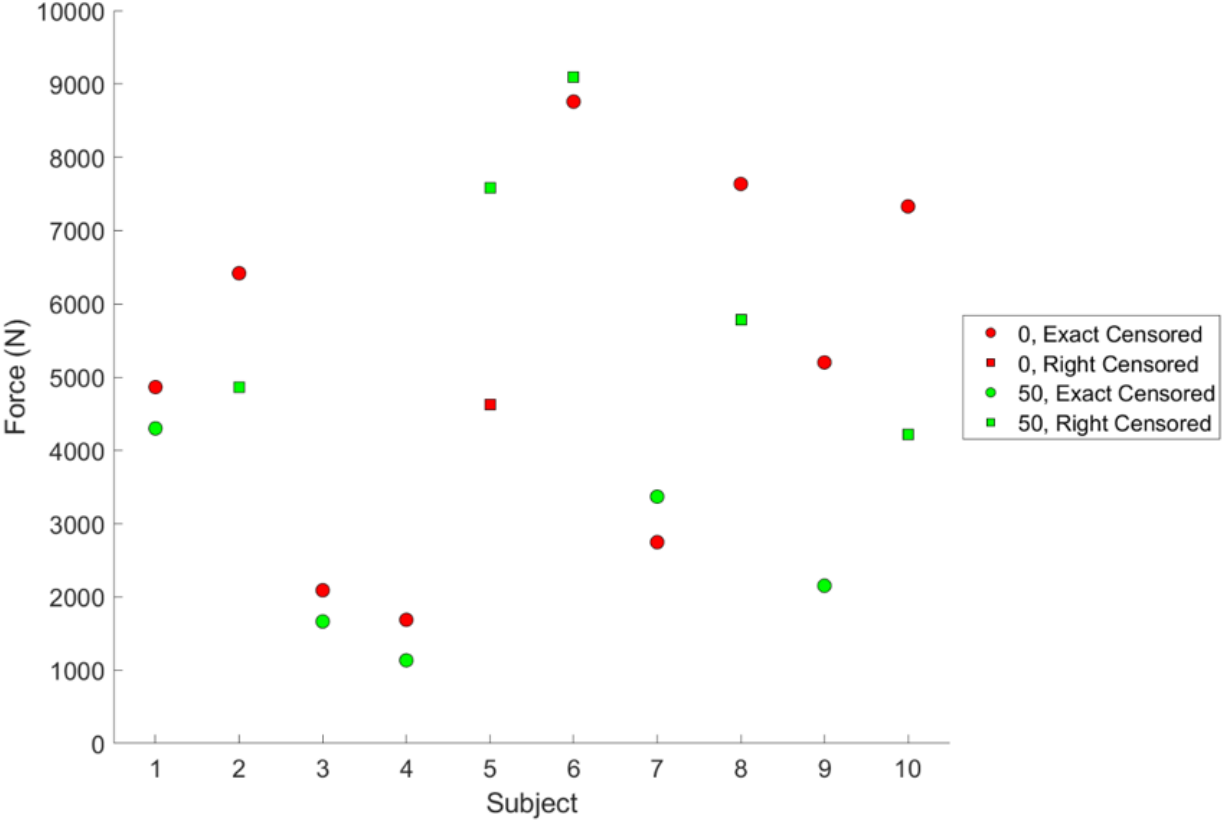